\documentstyle[amssymb,11pt]{article}

\newlength{\minitwocolumn}
\font\teneufm=eufm10
\font\seveneufm=eufm7
\font\fiveeufm=eufm5
\newfam\eufmfam
\textfont\eufmfam=\teneufm
\scriptfont\eufmfam=\seveneufm
\scriptscriptfont\eufmfam=\fiveeufm

\makeatletter
\@addtoreset{equation}{section}
\makeatother

\newtheorem{thm}{Theorem}[section]

\newtheorem{dfn}[thm]{Definition}
\title{
\Large{\bf
FREE FIELD REALIZATION OF\\
QUANTUM AFFINE SUPERALGEBRA
$U_q(\widehat{{sl}}(N|1))$
}}
\begin{document}
\maketitle
\begin{center}
{TAKEO KOJIMA}
\\~\\
{\it
Department of Mathematics and Physics,
Graduate School of Science and Engineering,\\
Yamagata University, Jonan 4-3-16, Yonezawa 992-8510,
Japan\\
kojima@yz.yamagata-u.ac.jp}
\end{center}

~\\
~\\

\begin{abstract}
We construct a free field realization of
the quantum affine superalgebra $U_q(\widehat{sl}(N|1))$
for an arbitrary level $k \in {\mathbb C}$.
\end{abstract}

~\\
~\\

\newpage

\section{Introduction}

The free field approach \cite{Jimbo-Miwa} provides a 
powerful method to
construct correlation functions of exactly solvable models.
In this paper we construct
a free field realization of the quantum affine superalgebra 
$U_q(\widehat{sl}(N|1))$ $(N\geqq 2)$
for an arbitrary level $k \in {\mathbb C}$.
The level parameter $k$ plays an important role 
in representation theory.
Free field realizations of an arbitrary level 
$k \in {\mathbb C}$
are completely different from those of level $k=1$.
In the case of level $k=1$,
free field realizations
\cite{Frenkel-Kac, Segal, Frenkel-Jing} 
have been constructed
for quantum affine algebra $U_q(g)$ 
in many cases
$g=(ADE)^{(r)}$ \cite{Frenkel-Jing, Jing1},
$(BC)^{(1)}$, $G_2^{(1)}$ \cite{Bernard,
Jing-Koyama-Misra, Jing2},
$\widehat{sl}(M|N)$, 
$osp(2|2)^{(2)}$
\cite{Kimura-Shiraishi-Uchiyama, Zhang, Yang-Zhang}.
In the case of an arbitrary level $k \in {\mathbb C}$,
free field realizations \cite{Wakimoto, Matsuo, Shiraishi}, 
have not yet been studied well
for quantum affine algebra $U_q(g)$.
In the case of an arbitrary level $k \in {\mathbb C}$,
free field realizations have been constructed
only for $U_q(\widehat{sl}(N))$ 
\cite{Awata-Odake-Shiraishi1}
and
$U_q(\widehat{sl}(2|1))$ \cite{Awata-Odake-Shiraishi2}.
The purpose of this paper is to construct
a free field realization of the quantum affine superalgebra 
$U_q(\widehat{sl}(N|1))$ for an arbitrary level 
$k \in {\mathbb C}$.
The representation theories of 
the superalgebra are much more complicated
than non-superalgebra and have rich structures
\cite{Kac1, Kac2,
Frappat-Sciarrino-Sorba, Kac-Wakimoto}.

This paper is organized as follows.
In section 2 we review 
the Chevalley realization of
the quantum superalgebra $U_q(sl(N|1))$ \cite{Yamane0}
and
the Drinfeld realization
of the quantum affine superalgebra 
$U_q(\widehat{sl}(N|1))$ \cite{Yamane}.
In section 3 
we review the Heisenberg realization of
quantum superalgebra $U_q(sl(N|1))$ \cite{Awata-Noumi-Odake}
and construct a free field realization of
the quantum affine superalgebra 
$U_q(\widehat{sl}(N|1))$ for an arbitrary level 
$k \in {\mathbb C}$.
In appendix A we 
explain how to find the free field realization of 
affine $U_q(\widehat{sl}(N|1))$ from the Heisenberg
realization $U_q(sl(N|1))$.
In appendix B we summarize some useful formulae.

\section{Quantum 
Affine Superalgebra $U_q(\widehat{sl}(N|1))$}

In this section we review 
the Chevalley realization of
the quantum superalgebra $U_q(sl(N|1))$
\cite{Yamane0} and
the Drinfeld realization of the quantum
superalgebra $U_q(\widehat{sl}(N|1))$
\cite{Yamane, Drinfeld} for $N=2, 3, 4,\cdots$.
We fix a complex number $q \neq 0, |q|<1$.
In what follows we use
\begin{eqnarray}
~[x,y]=xy-yx,\\
~\{x,y\}=xy+yx,\\
~[a]_q=\frac{q^a-q^{-a}}{
q-q^{-1}}.
\end{eqnarray}

\subsection{Quantum Superalgebra $U_q(sl(N|1))$}

Let us recall the definition of the quantum superalgebra
$U_q(sl(N|1))$ \cite{Yamane0}.
We set $\nu_1=\nu_2=\cdots=\nu_N=+, \nu_{N+1}=-$.
The Cartan matrix $(A_{i,j})_{1\leqq i,j \leqq N}$ 
of the Lie algebra 
${sl}(N|1)$ is given by
\begin{eqnarray}
A_{i,j}=
(\nu_i+\nu_{i+1})\delta_{i,j}-
\nu_i \delta_{i,j+1}-\nu_{i+1}\delta_{i+1,j}.
\end{eqnarray}
The diagonal part is
$(A_{i,i})_{1\leqq i \leqq N}
=(\overbrace{2, \cdots, 2}^{N-1},0)$.

\begin{dfn}~~\cite{Yamane0}~~
The Chevalley generators of
the quantum
superalgebra $U_q({sl}(N|1))$ are
\begin{eqnarray}
h_i, e_i, f_i~~~(1\leqq i \leqq N).
\end{eqnarray}
Defining relations are
\begin{eqnarray}
&&[h_i,h_j]=0,\\
&&[h_i,e_j]=A_{i,j}e_j,\\
&&[h_i,f_j]=-A_{i,j}f_j,\\
&&[e_i,f_j]=\delta_{i,j}\frac{q^{h_i}-q^{-h_i}}{q-q^{-1}}~~
{\rm for}~(i,j)\neq (N,N),\\
&&\{e_N,f_N\}=\frac{q^{h_N}-q^{-h_N}}{q-q^{-1}},
\end{eqnarray}
and the Serre relations
\begin{eqnarray}
&&e_ie_ie_j-(q+q^{-1})e_ie_je_i+
e_je_ie_i=0  ~~for~ |A_{i,j}|=1, i \neq N,\\
&&f_if_if_j-(q+q^{-1})f_if_jf_i+
f_jf_if_i=0  ~~for~ |A_{i,j}|=1, i \neq N.
\end{eqnarray}
\end{dfn}

\subsection{Quantum Affine Superalgebra $U_q(\widehat{sl}(N|1))$}

Let us recall
the definition of
the quantum affine superalgebra $U_q(\widehat{sl}(N|1))$
\cite{Yamane}.
The Cartan matrix 
$(A_{i,j})_{0\leqq i,j \leqq N}$
of the affine Lie algebra $\widehat{sl}(N|1)$ is
given by
\begin{eqnarray}
A_{i,j}=
(\nu_i+\nu_{i+1})\delta_{i,j}-
\nu_i \delta_{i,j+1}-\nu_{i+1}\delta_{i+1,j}.
\end{eqnarray}
Here we 
should read the suffixes $j$ of $\nu_j$
mod.$(N+1)$, i.e. $\nu_0=\nu_{N+1}$.
Here the diagonal part is
$(A_{i,i})_{0\leqq i \leqq N}
=(0,\overbrace{2, \cdots, 2}^{N-1},0)$.

\begin{dfn}~~\cite{Yamane}~
The Drinfeld generators of
the quantum affine
superalgebra $U_q(\widehat{sl}(N|1))$
are
\begin{eqnarray}
x_{i,m}^\pm,~h_{i,m},~c,~~(1\leqq i \leqq N,
m \in {\mathbb Z}).
\end{eqnarray}
Defining relations are
\begin{eqnarray}
&&~c : {\rm central},~[h_i,h_{j,m}]=0,\\
&&~[a_{i,m},h_{j,n}]=\frac{[A_{i,j}m]_q[cm]_q}{m}q^{-c|m|}
\delta_{m+n,0}~~(m,n\neq 0),\\
&&~[h_i,x_j^\pm(z)]=\pm A_{i,j}x_j^\pm(z),\\
&&~[h_{i,m}, x_j^+(z)]=\frac{[A_{i,j}m]_q}{m}
q^{-c|m|} z^m x_j^+(z)~~(m \neq 0),\\
&&~[h_{i,m}, x_j^-(z)]=-\frac{[A_{i,j}m]_q}{m}
z^m x_j^-(z)~~(m \neq 0),\\
&&(z_1-q^{\pm A_{i,j}}z_2)
x_i^\pm(z_1)x_j^\pm(z_2)
=
(q^{\pm A_{j,i}}z_1-z_2)
x_j^\pm(z_2)x_i^\pm(z_1)~~~{\rm for}~|A_{i,j}|\neq 0,
\\
&&
x_i^\pm(z_1)x_j^\pm(z_2)
=
x_j^\pm(z_2)x_i^\pm(z_1)~~~{\rm for}~|A_{i,j}|=0, (i,j)\neq (N,N),
\\
&&
\{x_N^\pm(z_1), x_N^\pm(z_2)\}=0,\label{def:Drinfeld8}\\
&&~[x_i^+(z_1),x_j^-(z_2)]
=\frac{\delta_{i,j}}{(q-q^{-1})z_1z_2}
\left(
\delta(q^{-c}z_1/z_2)\psi_i^+(q^{\frac{c}{2}}z_2)-
\delta(q^{c}z_1/z_2)\psi_i^-(q^{-\frac{c}{2}}z_2)
\right), \nonumber\\
&& ~~~~~{\rm for}~~(i,j) \neq (N,N),\\
&&~\{x_N^+(z_1),x_N^-(z_2)\}
=\frac{1}{(q-q^{-1})z_1z_2}
\left(
\delta(q^{-c}z_1/z_2)\psi_N^+(q^{\frac{c}{2}}z_2)-
\delta(q^{c}z_1/z_2)\psi_N^-(q^{-\frac{c}{2}}z_2)
\right), \nonumber\\
\\
&& 
\left(
x_i^\pm(z_{1})
x_i^\pm(z_{2})
x_j^\pm(z)-(q+q^{-1})
x_i^\pm(z_{1})
x_j^\pm(z)
x_i^\pm(z_{2})
+x_j^\pm(z)
x_i^\pm(z_{1})
x_i^\pm(z_{2})\right)\nonumber\\
&&+\left(z_1 \leftrightarrow z_2\right)=0
~~~{\rm for}~|A_{i,j}|=1,~i\neq N.
\end{eqnarray}
where we have used
$\delta(z)=\sum_{m \in {\mathbb Z}}z^m$.
Here we have 
used the abbreviation $h_i={h_{i,0}}$.
We have
set the generating function
\begin{eqnarray}
x_j^\pm(z)&=&
\sum_{m \in {\mathbb Z}}x_{j,m}^\pm z^{-m-1},\\
\psi_i^+(q^{\frac{c}{2}}z)&=&q^{h_i}
\exp\left(
(q-q^{-1})\sum_{m>0}h_{i,m}z^{-m}
\right),\\
\psi_i^-(q^{-\frac{c}{2}}z)&=&q^{-h_i}
\exp\left(-(q-q^{-1})\sum_{m>0}h_{i,-m}z^m\right).
\end{eqnarray}
\end{dfn}
We changed the gauge of boson $h_{i,m}$
from those of \cite{Yamane} and revised 
a misprint (\ref{def:Drinfeld8})
in \cite{Yamane}.

\section{Free Field Realization}

In this section we 
review the Heisenberg realization of $U_q(sl(N|1))$
\cite{Awata-Odake-Shiraishi2} and
construct a free field realization
of the quantum affine superalgebra $U_q(\widehat{sl}(N|1))$
for an arbitrary level $k \in {\mathbb C}$.

\subsection{Heisenberg Realization}

Let us recall the Heisenberg realization
of quantum superalgebra $U_q(sl(N|1))$
\cite{Awata-Odake-Shiraishi2}.
We introduce the coordinates 
$x_{i,j}$, $(1\leqq i<j \leqq N+1)$ by
\begin{eqnarray}
x_{i,j}=\left\{\begin{array}{cc}
z_{i,j}&~~~(1\leqq i<j \leqq N),\\
\theta_{i,j}&~~~(1\leqq i \leqq N, j=N+1).
\end{array}
\right.
\end{eqnarray}
Here $z_{i,j}$ are complex variables and
$\theta_{i,N+1}$ are the Grassmann odd variables
that satisfy
$\theta_{i,N+1}\theta_{i,N+1}=0$ and
$\theta_{i,N+1}\theta_{j,N+1}=-\theta_{j,N+1}\theta_{i,N+1}$,
$(i \neq j)$.
We introduce the differential operators
$\vartheta_{i,j}=x_{i,j}
\frac{\partial}{\partial x_{i,j}}$, 
$(1\leqq i<j \leqq N+1)$.
We fix parameters $\lambda_i \in {\mathbb C}$, 
$(1\leqq i \leqq N)$.
We set the differential operators
$H_i, E_i, F_i$, $(1\leqq i \leqq N)$ by
\begin{eqnarray}
H_i=\sum_{j=1}^{N}H_{i,j},~~~
E_i=\sum_{j=1}^i E_{i,j},~~~
F_i=\sum_{j=1}^N F_{i,j}.
\label{def:Heisenberg}
\end{eqnarray}
Here we have set
\begin{eqnarray}
H_{i,j}
&=&
\left\{
\begin{array}{cc}
\nu_i \vartheta_{j,i}-\nu_{i+1}\vartheta_{j,i+1}&~~(1\leqq j \leqq i-1),\\
\lambda_i-(\nu_i+\nu_{i+1})\vartheta_{i,i+1}&~~(j=i),\\
\nu_{i+1}\vartheta_{i+1,j+1}-\nu_i \vartheta_{i,j+1}&~~(i+1 \leqq j \leqq N),
\end{array}
\right.
\end{eqnarray}
\begin{eqnarray}
E_{i,j}&=&
\frac{x_{j,i}}{x_{j,i+1}}[\vartheta_{j,i+1}]_q
~q^{
\sum_{l=1}^{j-1}(\nu_i \vartheta_{l,i}-\nu_{i+1}
\vartheta_{l,i+1})},
\end{eqnarray}
\begin{eqnarray}
F_{i,j}&=&
\left\{\begin{array}{cc}
\begin{array}{c}\nu_i \frac{x_{j,i+1}}{x_{j,i}}[
\vartheta_{j,i}]_q \times \\
\times~q^{\sum_{l=j+1}^{i-1}
(\nu_{i+1}\vartheta_{l,i+1}-\nu_i \vartheta_{l,i})
-\lambda_i+(\nu_i+\nu_{i+1})\vartheta_{i,i+1}
+\sum_{l=i+2}^{N+1}(\nu_i \vartheta_{i,l}-\nu_{i+1}
\vartheta_{i+1,l})}
\end{array}
&~~(1 \leqq j \leqq i-1),\\
x_{i,i+1}\left[\lambda_i-\nu_i \vartheta_{i,i+1}-
\sum_{l=i+2}^{N+1}
(\nu_i \vartheta_{i,l}-\nu_{i+1} \vartheta_{i+1,l})
\right]_q&~~(j=i),\\
-\nu_{i+1}\frac{x_{i,j+1}}{x_{i+1,j+1}}
[\vartheta_{i+1,j+1}]_q q^{
\lambda_i+\sum_{l=j+1}^{N+1}
(\nu_{i+1}\vartheta_{i+1,l}-\nu_i \vartheta_{i,l})}&~~
(i+1 \leqq j \leqq N).
\end{array}
\right.\nonumber\\
\end{eqnarray}
Here we read $x_{i,i}=1$ and, for Grassmann odd variables
$x_{i,j}$, the expression $\frac{1}{x_{i,j}}$
stands for the derivative $\frac{1}{x_{i,j}}=
\frac{\partial}{\partial x_{i,j}}$.

\begin{thm}~~\cite{Awata-Odake-Shiraishi2}~~
A Heisenberg realization of the quantum superalgebra
$U_q(sl(N|1))$ is given in the following way.
\begin{eqnarray}
h_i&\to& H_i,\\
e_i&\to& E_i,\\
f_i&\to& F_i.
\end{eqnarray}
\end{thm}

In appendix \ref{appendixA}
we explain how to find the free field realization of 
affine $U_q(\widehat{sl}(N|1))$ from this Heisenberg
realization $U_q(sl(N|1))$.

\subsection{Boson}

Let us fix the level $c=k \in {\mathbb C}$.
Let us introduce the bosons
and the zero-mode operators
$a_m^j, Q_a^j$ $(m \in {\mathbb Z},
1\leqq j \leqq N)$, 
$b_m^{i,j}, Q_b^{i,j}$,
$c_m^{i,j}, Q_c^{i,j}$
$(m \in {\mathbb Z}, 1\leqq i<j \leqq N+1)$.
The bosons $a_m^i, b_m^{i,j}, c_m^{i,j}$, 
$(m \in {\mathbb Z}_{\neq 0})$ satisfy
\begin{eqnarray}
&&~[a_m^i,a_n^j]=\frac{[(k+N-1)m]_q[A_{i,j}m]_q}{m}
\delta_{m+n,0},
\\
&&~[b_m^{i,j},b_n^{i',j'}]=
-\nu_i \nu_j \frac{[m]_q^2}{m}
\delta_{i,i'}\delta_{j,j'}\delta_{m+n,0},
\\
&&~[c_m^{i,j},c_n^{i',j'}]=
\nu_i \nu_j \frac{[m]_q^2}{m}
\delta_{i,i'}\delta_{j,j'}
\delta_{m+n,0}.
\end{eqnarray}
The zero-mode operators $a_0^i,Q_a^i$, $b_0^{i,j},Q_b^{i,j}$,
$c_0^{i,j}, Q_c^{i,j}$ satisfy
\begin{eqnarray}
&&[a_0^i, Q_a^j]=(k+N-1)A_{i,j}, 
\\
&&[b_0^{i,j},Q_b^{i',j'}]=
-\nu_i \nu_j \delta_{i,i'}\delta_{j,j'},
\\
&&[c_0^{i,j},Q_c^{i',j'}]=
\nu_i \nu_j \delta_{i,i'}\delta_{j,j'}.
\end{eqnarray}
and other commutators vanish.
We impose the cocycle condition on 
the zero-mode operator $Q_{b}^{i,j}$, $(1\leqq i<j \leqq N+1)$ by
\begin{eqnarray}
~[Q_b^{i,j},Q_b^{i',j'}]=\delta_{j,N+1}\delta_{j',N+1}
\pi \sqrt{-1}~~~~~{\rm for}~(i,j) \neq (i',j').
\end{eqnarray}
We have the following (anti)commutation relations
\begin{eqnarray}
&&
\left[\exp\left(Q_b^{i,j}\right),\exp\left(Q_b^{i',j'}\right)
\right]=0
~~~
(1\leqq i<j \leqq N, 1\leqq i'<j' \leqq N),\\
&&\left\{\exp\left(Q_b^{i,N+1}\right),\exp\left(Q_b^{j,N+1}\right)
\right\}=0~~~
(1\leqq i \neq j \leqq N).
\end{eqnarray}
We use the following normal ordering symbol $: :$ as follows.
\begin{eqnarray}
&&:b_m^{i,j}b_n^{i',j'}:=
\left\{
\begin{array}{cc}
b_m^{i,j}b_n^{i',j'}&~~(m<0),\\
b_n^{i',j'}b_m^{i,j}&~~(m>0),
\end{array}\right.
~~:a_m^{i}a_n^{j}:=
\left\{
\begin{array}{cc}
a_m^{i}a_n^{j}&~~(m<0),\\
a_n^{j}a_m^{i}&~~(m>0),
\end{array}\right.\\
&&:b_0^{i,j} Q_b^{i',j'}:=
:Q_b^{i',j'} b_0^{i,j}:=
Q_b^{i',j'} b_0^{i,j},~~
:a_0^{i} Q_a^{j}:=:Q_a^{j} a_0^{i}:=
Q_a^{j} a_0^{i}.
\end{eqnarray}
The above boson structure is the straightforward generalization
of those in \cite{Awata-Odake-Shiraishi2}.
Note that $(N-1)$ is the dual Coxter number.
In what follows we use 
$\{a_m^j (1\leqq j \leqq N),
b_m^{i,j}, Q_b^{i,j} (1\leqq i<j \leqq N+1),
c_m^{i,j}, Q_c^{i,j} (1\leqq i<j \leqq N)\}$
which is a subset of the above boson system.
In what follows we use the abbreviations $b^{i,j}(z), c^{i,j}(z), b_\pm^{i,j}(z),
a^j_\pm(z)$.
\begin{eqnarray}
&&b^{i,j}(z)=
-\sum_{m \neq 0}\frac{b_m^{i,j}}{[m]_q}z^{-m}+Q_b^{i,j}+b_0^{i,j}{\rm log}z,
\\
&&c^{i,j}(z)=
-\sum_{m \neq 0}\frac{c_m^{i,j}}{[m]_q}z^{-m}+Q_c^{i,j}+c_0^{i,j}{\rm log}z,
\\
&&b_\pm^{i,j}(z)=\pm (q-q^{-1})\sum_{\pm m>0}b_m^{i,j} 
z^{-m} \pm b_0^{i,j}{\rm log}q,\\
&&a_\pm^{j}(z)=\pm (q-q^{-1})\sum_{\pm m>0}a_m^{j} 
z^{-m}\pm a_0^j {\rm log}q.
\end{eqnarray}

\subsection{Free Field Realization}

In this section
we construct a free field realization of the quantum 
affine superalgebra
$U_q(\widehat{sl}(N|1))$ for an arbitrary level $k$.
In \cite{Awata-Noumi-Odake}, 
on the basis of the Heisenberg realization of
the quantum algebra $U_q(sl(N))$, a free field realization of
the quantum affine algebra $U_q(\widehat{sl}(N))$
was obtained.
Here we try to generalize it to the quantum
affine superalgebra $U_q(\widehat{sl}(N|1))$.
Detailed calculations of this trial are summarized in 
appendix \ref{appendixA}.
We introduce the operators
$X_i^\pm(z), \Psi_i^\pm(z)$, $(1\leqq i \leqq N)$ 
on the Fock space as follows.
For $1\leqq i \leqq N-1$ we introduce
\begin{eqnarray}
X_i^+(z)&=&\frac{1}{(q-q^{-1})z}\sum_{j=1}^i(X_{i,2j-1}^+(z)-X_{i,2j}^+(z)),
\label{boson1}
\\
X_N^+(z)&=&\sum_{j=1}^N X_{N,j}^+(z),
\label{boson2}
\\
X_i^-(z)&=&\frac{1}{(q-q^{-1})z}\left(
\sum_{j=1}^{i-1}(X_{i,2j-1}^-(z)-X_{i,2j}^-(z))
+(X_{i,2i-1}^-(z)-X_{i,2i}^-(z))
\right.
\nonumber\\
&&\left.-
\sum_{j=i+1}^{N-1}(X_{i,2j-1}^-(z)-X_{i,2j}^-(z))\right)
+q^{k+N-1}X_{i,2N-1}^-(z),
\label{boson3}
\\
X_N^-(z)&=&\frac{1}{(q-q^{-1})z}\sum_{j=1}^N
\left(-q^{j-1}X_{N,2j-1}^-(z)+q^{j-1}X_{N,2j}^-(z)\right).
\label{boson4}
\end{eqnarray}
\begin{eqnarray}
\Psi_i^\pm(q^{\pm \frac{k}{2}}z)&=&
\exp\left(
a_\pm^i(q^{\pm \frac{k+N-1}{2}}z)+
\sum_{l=1}^i(b_\pm^{l,i+1}(q^{\pm(l+k-1)}z)-b_\pm^{l,i}
(q^{\pm(l+k)}z))
\right.\nonumber
\\
&&
+\sum_{l=i+1}^{N}(b_\pm^{i,l}(q^{\pm(k+l)}z)-
b_\pm^{i-1,l}(q^{\pm(k+l-1)}z))\nonumber\\
&&
\left.
+b_\pm^{i,N+1}(q^{\pm(k+N)}z)-
b_\pm^{i+1,N+1}(q^{\pm(k+N-1)}z)\right),
\label{boson5}
\\
\Psi_N^\pm(q^{\pm \frac{k}{2}}z)&=&
\exp\left(a_\pm^N(q^{\pm \frac{k+N-1}{2}}z)-
\sum_{l=1}^{N-1}
(b_\pm^{l,N}(q^{\pm (k+l)}z)
+b_\pm^{l,N+1}(q^{\pm (k+l)}z))\right).
\label{boson6}
\end{eqnarray}
Here we have used
the auxiliary bosonic operators $X_{i,j}^\pm(z)$ as follows.\\
For $1\leqq i \leqq N-1$ and $1\leqq j \leqq i$ we set
\begin{eqnarray}
X_{i,2j-1}^+(z)&=&
:\exp\left((b+c)^{j,i}(q^{j-1}z)+b_+^{j,i+1}(q^{j-1}z)-
(b+c)^{j,i+1}(q^jz)\right.\nonumber\\
&&\left.+\sum_{l=1}^{j-1}
(b_+^{l,i+1}(q^{l-1}z)-b_+^{l,i}(q^lz))\right):,
\label{boson7}\\
X_{i,2j}^+(z)&=&
:\exp\left((b+c)^{j,i}(q^{j-1}z)+b_-^{j,i+1}(q^{j-1}z)-
(b+c)^{j,i+1}(q^{j-2}z)\right.\nonumber\\
&&\left.+\sum_{l=1}^{j-1}
(b_+^{l,i+1}(q^{l-1}z)-b_+^{l,i}(q^lz))\right):.
\label{boson8}
\end{eqnarray}
For $1\leqq j \leqq N$ we set
\begin{eqnarray}
X_{N,j}^+(z)&=&:\exp\left(
(b+c)^{j,N}(q^{j-1}z)
+b^{j,N+1}(q^{j-1}z)\right.\nonumber\\
&&\left.-\sum_{l=1}^{j-1}(b_+^{l,N+1}(q^lz)+b_+^{l,N}(q^lz))
\right):.\label{boson9}
\end{eqnarray}
For $1\leqq i \leqq N-1$ and $1\leqq j \leqq i-1$ we set
\begin{eqnarray}
X_{i,2j-1}^-(z)&=&
:\exp\left(
a_-^i(q^{-\frac{k+N-1}{2}}z)
+(b+c)^{j,i+1}(q^{-k-j}z)\right.\nonumber\\
&&-b_-^{j,i}(q^{-k-j}z)
-(b+c)^{j,i}(q^{-k-j+1}z)\nonumber\\
&& 
+\sum_{l=j+1}^i 
(b_-^{l,i+1}(q^{-k-l+1}z)-b_-^{l,i}(q^{-k-l}z))\nonumber\\
&&
+\sum_{l=i+1}^N
(b_-^{i,l}(q^{-k-l}z)-b_-^{i+1,l}(q^{-k-l+1}z))\nonumber\\
&&
\left.+b_-^{i,N+1}(q^{-k-N}z)-b_-^{i+1,N+1}(q^{-k-N+1}z)
\right):,
\label{boson10}
\\
X_{i,2j}^-(z)&=&
:\exp\left(a_-^i(q^{-\frac{k+N-1}{2}}z)
+(b+c)^{j,i+1}(q^{-k-j}z)\right.\nonumber\\
&&-b_+^{j,i}(q^{-k-j}z)
-(b+c)^{j,i}(q^{-k-j-1}z)\nonumber\\
&& 
+\sum_{l=j+1}^i 
(b_-^{l,i+1}(q^{-k-l+1}z)-b_-^{l,i}(q^{-k-l}z))
\nonumber\\
&&
+\sum_{l=i+1}^N
(b_-^{i,l}(q^{-k-l}z)-b_-^{i+1,l}(q^{-k-l+1}z))\nonumber\\
&&
\left.+b_-^{i,N+1}(q^{-k-N}z)-b_-^{i+1,N+1}(q^{-k-N+1}z)\right):.
\label{boson11}
\end{eqnarray}
For $1\leqq i \leqq N-1$ we set
\begin{eqnarray}
X_{i,2i-1}^-(z)&=&:\exp\left(a_-^i(q^{-\frac{k+N-1}{2}}z)
+(b+c)^{i,i+1}(q^{-k-i}z)\right.\nonumber\\
&&+\sum_{l=i+1}^N(b_-^{i,l}(q^{-k-l}z)
-b_-^{i+1,l}(q^{-k-l+1}z))\nonumber\\
&&
\left.
+b_-^{i,N+1}(q^{-k-N}z)-b_-^{i+1,N+1}(q^{-k-N+1}z)\right):,
\label{boson12}
\\
X_{i,2i}^-(z)&=&:
\exp\left(a_+^i(q^{\frac{k+N-1}{2}}z)
+(b+c)^{i,i+1}(q^{k+i}z)\right.\nonumber\\
&&+\sum_{l=i+1}^N(b_+^{i,l}(q^{k+l}z)
-b_+^{i+1,l}(q^{k+l-1}z))\nonumber\\
&&
\left.
+b_+^{i,N+1}(q^{k+N}z)-b_+^{i+1,N+1}(q^{k+N-1}z)\right):.
\label{boson13}
\end{eqnarray}
For $1\leqq i \leqq N-1$ and $i+1 \leqq j \leqq N-1$ we set
\begin{eqnarray}
X_{i,2j-1}^-(z)&=&
:\exp\left(a_+^i(q^{\frac{k+N-1}{2}}z)
+(b+c)^{i,j+1}(q^{k+j}z)\right.\nonumber\\
&&+b_+^{i+1,j+1}(q^{k+j}z)-(b+c)^{i+1,j+1}(q^{k+j+1}z)
\nonumber\\
&&
+\sum_{l=j+1}^N
(b_+^{i,l}(q^{k+l}z)-b_+^{i+1,l}(q^{k+l-1}z))\nonumber\\
&&\left.+b_+^{i,N+1}(q^{k+N}z)-b_+^{i+1,N+1}(q^{k+N-1}z)\right):,
\label{boson14}
\\
X_{i,2j}^-(z)&=&
:\exp\left(a_+^i(q^{\frac{k+N-1}{2}}z)
+(b+c)^{i,j+1}(q^{k+j}z)\right.\nonumber\\
&&+b_-^{i+1,j+1}(q^{k+j}z)-(b+c)^{i+1,j+1}(q^{k+j-1}z)
\nonumber\\
&&
+\sum_{l=j+1}^N
(b_+^{i,l}(q^{k+l}z)-b_+^{i+1,l}(q^{k+l-1}z))\nonumber\\
&&\left.
+b_+^{i,N+1}(q^{k+N}z)-b_+^{i+1,N+1}(q^{k+N-1}z)\right):.
\label{boson15}
\end{eqnarray}
For $1\leqq i \leqq N-1$ we set
\begin{eqnarray}
X_{i,2N-1}^-(z)&=&:\exp\left(a_+^i(q^{\frac{k+N-1}{2}}z)
-b^{i,N+1}(q^{k+N-1}z)\right.\nonumber\\
&&\left.-b_+^{i+1,N+1}(q^{k+N-1}z)+b^{i+1,N+1}(q^{k+N}z)
\right):.
\label{boson16}
\end{eqnarray}
For $1\leqq j \leqq N-1$ we set
\begin{eqnarray}
X_{N,2j-1}^-(z)&=&:
\exp\left(a_-^N(q^{-\frac{k+N-1}{2}}z)
-b_-^{j,N}(q^{-k-j}z)-(b+c)^{j,N}(q^{-k-j+1}z)\right.\nonumber\\
&&
-b_-^{j,N+1}(q^{-k-j}z)-b^{j,N+1}(q^{-k-j+1}z)\nonumber\\
&&\left.-\sum_{l=j+1}^{N-1}
(b_-^{l,N}(q^{-k-l}z)+b_-^{l,N+1}(q^{-k-l}z))\right):,
\label{boson17}
\\
X_{N,2j}^-(z)&=&:\exp\left(
a_-^N(q^{-\frac{k+N-1}{2}}z)
-b_+^{j,N}(q^{-k-j}z)-(b+c)^{j,N}(q^{-k-j-1}z)\right.
\nonumber\\
&&-b_+^{j,N+1}(q^{-k-j}z)-b^{j,N+1}(q^{-k-j-1}z)\nonumber\\
&&\left.
-\sum_{l=j+1}^{N-1}
(b_-^{l,N}(q^{-k-l}z)+b_-^{l,N+1}(q^{-k-l}z))\right):,
\label{boson18}
\\
X_{N,2N-1}^-(z)&=&
:\exp\left(
a_-^N(q^{-\frac{k+N-1}{2}}z)-b^{N,N+1}(q^{-k-N+1}z)\right):,
\label{boson19}
\\
X_{N,2N}^-(z)&=&:
\exp\left(
a_+^N(q^{\frac{k+N-1}{2}}z)-b^{N,N+1}(q^{k+N-1}z)\right):.
\label{boson20}
\end{eqnarray}
Now we have introduced the bosonic operators 
$X_i^\pm(z)$ and $\Psi_i^\pm(z)$.\\
The following is {\bf main result} of this paper.

\begin{thm}~~
A free field realization
of the quantum affine superalgebra
$U_q(\widehat{sl}(N|1))$ is given in the following way.
\begin{eqnarray}
c &\mapsto& k \label{thm:1}\\
x_i^\pm(z) &\mapsto& X_i^\pm(z) \label{thm:2}\\
\psi_i^\pm(z) &\mapsto& \Psi_i^\pm(z) \label{thm:3}. 
\end{eqnarray}
\end{thm}

~\\
In other words, the above map gives a homomorphism from
$U_q(\widehat{sl}(N|1))$ to the bosonic operator.
Very explicitly the relation (\ref{thm:3}) is written as
\begin{eqnarray}
h_{i,m} &\mapsto&
q^{-\frac{k+N-1}{2}|m|}a_m^i+
\sum_{l=1}^i(q^{-(k+l-1)|m|}b_m^{l,i+1}
-q^{-(k+l)|m|}b_m^{l,i})\nonumber
\\
&&+
\sum_{l=i+1}^N(q^{-(k+l)|m|}b_m^{i,l}-
q^{-(k+l-1)|m|}b_m^{i+1,l})\nonumber\\
&&+q^{-(k+N)|m|}b_m^{i,N+1}-
q^{-(k+N-1)|m|}b_m^{i+1,N+1}~~(1\leqq i \leqq N-1),\\
h_{N,m}
&\mapsto&
q^{-\frac{k+N-1}{2}|m|}a_m^N-
\sum_{l=1}^{N-1}(q^{-(k+l)|m|}b_m^{l,N}+q^{-(k+l)|m|}
b_m^{l,N+1}).
\end{eqnarray}

We give some comments on this realization.
Upon the specialization $N=2$,
this free field realization 
reproduces the result for 
$U_q(\widehat{sl}(2|1))$ in
\cite{Awata-Odake-Shiraishi2}.
The structure of non-superalgebra $U_q(\widehat{sl}(N))$ exists inside
the superalgebra $U_q(\widehat{sl}(N|1))$.
Hence the free field realizations of the currents 
$X_i^\pm(z)$ $(i \neq N)$
for $U_q(\widehat{sl}(N|1))$ 
are quite similar as those for
$U_q(\widehat{sl}(N))$.
The free field realizations of the fermionic operators 
$X_{N,j}^+(z)$,
$X_{N,2j-1}^-(z), X_{N,2j}^-(z)$ and $X_{j,2N-1}^-(z)$ of
$U_q(\widehat{sl}(N|1))$ are
completely different from those of $U_q(\widehat{sl}(N))$.
The free field realization 
of this paper is not irreducible representation.
We have to construct screening currents that commute
with the currents $X_j^\pm(z)$ in order to
get an irreducible representation
\cite{Bernard-Felder, Konno, Zhang-Gould}.
We would like report this subject in the future publication.
Applying the dressing method developed in \cite{Kojima} to this theorem,
we have a free field realization of the elliptic algebra $U_{q,p}(\widehat{sl}(N|1))$.
\\
\\
{\it Proof of Theorem.}~~
Direct calculations of the normal orderings 
show this theorem.
The normal orderings
of bosonic operators $X_{i,j}^\pm(z)$ 
$(i \neq N, j \neq 2N-1)$ of
the superalgebra $U_q(\widehat{sl}(N|1))$
are exactly the same as those of the non-superalgebra
$U_q(\widehat{sl}(N))$.
Hence the proof of the relations
for the bosonic operators $X_{i}^\pm(z)$ 
$(i\neq N)$
is exactly the same as those of 
$U_q(\widehat{sl}(N))$.
Let us focus our attention on
the fermionic operators $X_N^\pm(z)$
that is new for the superalgebra.
We show the following relations
for the fermionic operators $X_N^\pm(z)$.
\begin{eqnarray}
&&~\{X_N^+(z_1),X_N^-(z_2)\}\nonumber\\
&&=\frac{1}{(q-q^{-1})z_1z_2}
\left(
\delta(q^{k}z_2/z_1)\Psi_N^+(q^{\frac{k}{2}}z_2)-
\delta(q^{-k}z_2/z_1)\Psi_N^-(q^{-\frac{k}{2}}z_2)
\right),
\label{thm:eqn1}
\end{eqnarray}
and
\begin{eqnarray}
&&~[X_N^+(z_1),X_j^-(z_2)]=0~~~~~
{\rm for}~1\leqq j \leqq N-1.
\label{thm:eqn2}
\end{eqnarray}

First, let us show (\ref{thm:eqn1}). 
Using the relation (\ref{eqn:a5}) in appendix \ref{appendixB},
we have
\begin{eqnarray}
&&\{X_N^+(z_1),X_N^-(z_2)\}\nonumber\\
&&=\frac{1}{(q-q^{-1})z_2}
\sum_{j=1}^N
q^{j-1}\left(-\{X_{N,j}^+(z_1),X_{N,2j-1}^-(z_2)\}
+\{X_{N,j}^+(z_1),X_{N,2j}^-(z_2)\}\right).
\nonumber
\end{eqnarray}
Using the relations
(\ref{eqn:a1}), (\ref{eqn:a2}), (\ref{eqn:a3})
and (\ref{eqn:a4}) in appendix \ref{appendixB}, 
we have
\begin{eqnarray}
&&\{X_N^+(z_1),X_N^-(z_2)\}=\frac{1}{(q-q^{-1})z_1z_2}
\left(
\delta(q^kz_2/z_1)\Psi_N^+(q^{\frac{k}{2}}z_2)
-\delta(q^{-k}z_2/z_1)\Psi_N^-(q^{-\frac{k}{2}}z_2)
\right)
\nonumber\\
&&
+
\frac{1}{(q-q^{-1})z_1z_2}
\exp\left(
a_-^N(q^{-\frac{k+N-1}{2}}z_2)\right)
\nonumber
\times
\\
&&
\left\{
\sum_{j=1}^{N-1}
\delta\left(\frac{q^{-k-2j}z_2}{z_1}\right)
:\exp
(
-\sum_{l=1}^{j}
(b_+^{l,N}(q^{l}z_1)+b_+^{l,N+1}(q^{l}z_1))
-\sum_{l=j+1}^{N-1}
(b_-^{l,N}(q^{-k-l}z_2)+b_+^{l,N+1}(q^{-k-l}z_2))
):\right.
\nonumber\\
&&\left.
-\sum_{j=2}^{N}
\delta\left(\frac{q^{-k-2j+2}z_2}{z_1}\right)
:\exp
(
-\sum_{l=1}^{j-1}(b_+^{l,N}(q^{l}z_1)+b_+^{l,N+1}(q^{l}z_1)
-\sum_{l=j}^{N-1}
(b_-^{l,N}(q^{-k-l}z_2)+b_+^{l,N+1}(q^{-k-l}z_2))
):\right\}.
\nonumber
\end{eqnarray}
Making the transformation $j \to j-1$
in the first sum $\sum_{j=1}^{N-1}\delta(q^{-k-2j}z_2/z_1)$,
we see cancellations.
We have the relation (\ref{thm:eqn1}).

Next, let us show (\ref{thm:eqn2}).
Using the relation (\ref{eqn:a9}) in appendix \ref{appendixB},
we have
the following for $1\leqq j \leqq N-2$.
\begin{eqnarray}
&&\left[X_N^+(z_1),X_j^-(z_2)\right]
\nonumber\\
&&=
\frac{-1}{(q-q^{-1})z_2}
\left[
X_{N,j}^+(z_1),X_{j,2N-3}^-(z_2)
\right]
+q^{k+N-1}
\left[
X_{N,j+1}^+(z_1),X_{j,2N-1}^-(z_2)\right].
\nonumber
\end{eqnarray}
Using the relations (\ref{eqn:a6}), (\ref{eqn:a8})
in appendix \ref{appendixB}, we have
\begin{eqnarray}
&&
\left[X_N^+(z_1),X_j^-(z_2)\right]=
\delta\left(\frac{q^{k+N-j}z_2}{z_1}\right)
\left(-\frac{1}{z_2}+\frac{q^{k+N-j}}{z_1}\right)
\nonumber\\
&&\times
:\exp\left(a_+^{j}(q^{\frac{k+N-1}{2}}z_2)
-b_+^{j+1,N+1}(q^{k+N-1}z_2)
+b^{j+1,N+1}(q^{k+N}z_2)
+(b+c)^{j,N}(q^{k+N-1}z_2)\right.
\nonumber\\
&&\left.-\sum_{l=1}^{j-1}
(b_+^{l,N}(q^{k+N-j+l}z_2)
+b_+^{l,N+1}(q^{k+N-j+l}z_2))\right):.\nonumber
\end{eqnarray}
From the relation
$\left(-\frac{1}{z_2}+\frac{q^{k+N-j}}{z_1}\right)
\delta\left(\frac{q^{k+N-j}z_2}{z_1}\right)=0$, we have
\begin{eqnarray}
\left[X_N^+(z_1),X_j^-(z_2)\right]=0~~~
{\rm for}~1\leqq j \leqq N-2.\nonumber
\end{eqnarray}
From the relation (\ref{eqn:a9}) in appendix \ref{appendixB},
we have
\begin{eqnarray}
&&\left[X_N^+(z_1),X_{N-1}^-(z_2)\right]\nonumber\\
&&=
\frac{-1}{(q-q^{-1})z_2}
\left[
X_{N,N-1}^+(z_1),X_{N-1,2N-2}^-(z_2)\right]
+q^{k+N-1}
\left[X_{N,N}^+(z_1),X_{N-1,2N-1}^-(z_2)\right].
\nonumber
\end{eqnarray}
Using the relations (\ref{eqn:a7}), (\ref{eqn:a8})
and
the relation $\delta\left(\frac{q^{k-1}z_2}{z_1}\right)
\left(-\frac{1}{z_2}+\frac{q^{k-1}}{z_1}\right)=0$,
we have
\begin{eqnarray}
&&\left[X_N^+(z_1),X_{N-1}^-(z_2)\right]=
\delta\left(\frac{q^{k-1}z_2}{z_1}\right)
\left(-\frac{1}{z_2}+\frac{q^{k-1}}{z_1}\right)
\nonumber\\
&&\times
:\exp\left(a_+^{N-1}(q^{\frac{k+N-1}{2}}z_2)
-b_+^{N,N+1}(q^{k+N-1}z_2)
+b^{N,N+1}(q^{k+N}z_2)
+(b+c)^{N-1,N}(q^{k+N-1}z_2)\right.
\nonumber\\
&&\left.-\sum_{l=1}^{N-2}
(b_+^{l,N}(q^{k+l+1}z_2)+b_+^{l,N+1}(q^{k+l+1}z_2))\right):
=0.\nonumber
\end{eqnarray}
We have shown the relation (\ref{thm:eqn2}).
~~~~~{Q.E.D.}

~\\

%%%%%%%%%%%%%%%%%%%%%%%%%%%%%%%%%%%%%%%

\subsection*{Acknowledgements}~~
This work is supported by the Grant-in-Aid for
Scientific Research {\bf C} (21540228)
from Japan Society for Promotion of Science. 
The author would like to thank
Professor Hiroyuki Yamane for informing
the author of a misprint in the paper \cite{Yamane}.
The author would like to thank
Professors Laszlo Feher, Hitoshi Konno and Akihiro Tsuchiya
for their interests to this work.
The author is grateful to
Professor Pascal Baseilhac 
and the colleagues in University of Tours
for kind invitation and warm hospitality
during his stay in Tours.
This paper is
dedicated to Professor Michio Jimbo 
on the occasion of his 60th birthday.

~\\

\begin{appendix}

\section{Replacement}
\label{appendixA}

In this appendix 
we explain how to find the free field realization of 
affine $U_q(\widehat{sl}(N|1))$ from the Heisenberg
realization of $U_q(sl(N|1))$.

\subsection{Basic Operator}

We would like to explain the role of
the basic operators 
\begin{eqnarray}
:\exp\left(\pm b^{i,N+1}(z)\right):,~ 
:\exp\left(b_\pm^{i,j}(z)\pm (b+c)^{i,j}(q^{\mp 1}z)\right):,
\label{def:basic}
\end{eqnarray}
which have been used for $U_q(\widehat{sl}(2|1))$
\cite{Awata-Odake-Shiraishi2} and 
$U_q(\widehat{sl}(2))$ \cite{Shiraishi},
respectively.
The basic operators $:\exp\left(\pm b^{i,N+1}(z)\right):$
$(1\leqq i \leqq N)$ satisfy the fermionic relation
\begin{eqnarray}
\left\{:\exp(b^{i,N+1}(z_1)):,
:\exp(-b^{i,N+1}(z_2)):
\right\}
=\frac{1}{z_1}\delta(z_2/z_1).
\end{eqnarray}
The basic operators 
$:\exp\left(\pm b^{i,N+1}(z)\right):$
create the delta-function $\delta(z)$ and
play important roles in constructions
of the fermionic operators $X_N^\pm(z)$ 
that satisfy
\begin{eqnarray}
~\{X_N^+(z_1),X_N^-(z_2)\}
=\frac{1}{(q-q^{-1})z_1z_2}
\left(
\delta(q^{k}z_2/z_1)\Psi_N^+(q^{\frac{k}{2}}z_2)-
\delta(q^{-k}z_2/z_1)\Psi_N^-(q^{-\frac{k}{2}}z_2)
\right).\nonumber
\end{eqnarray}
The basic
operators
$:\exp\left(b_\pm^{i,j}(z)\pm (b+c)^{i,j}(q^{\mp 1}z)\right):$
$(1\leqq i<j\leqq N)$
satisfy the bosonic relations
\begin{eqnarray}
&&\left[
:\exp\left(b_+^{i,j}(z_1)-(b+c)^{i,j}(qz_1)\right):,
:\exp\left(b_+^{i,j}(z_2)+(b+c)^{i,j}(q^{-1}z_2)\right):
\right]\nonumber\\
&&=(q^{-1}-q)\delta(q^{-2}z_2/z_1):\exp\left(b_+^{i,j}(z_1)+b_+^{i,j}(z_2)\right):,\\
&&\left[
:\exp\left(b_-^{i,j}(z_1)-(b+c)^{i,j}(q^{-1}z_1)\right):,
:\exp\left(b_-^{i,j}(z_2)+(b+c)^{i,j}(qz_2)\right):
\right]\nonumber\\
&&=
(q-q^{-1})\delta(q^2z_2/z_1):\exp\left(
b_-^{i,j}(z_1)+b_-^{i,j}(z_2)\right):.
%\\
%&&
%\left[
%\exp\left(b_+^{i,j}(z_1)-(b+c)^{i,j}(qz_1)\right),
%\exp\left(b_-^{i,j}(z_2)+(b+c)^{i,j}(qz_2)\right)
%\right]=0,\\
%&&\left[
%\exp\left(b_-^{i,j}(z_1)-(b+c)^{i,j}(q^{-1}z_1)\right),
%\exp\left(b_+^{i,j}(z_2)+(b+c)^{i,j}(q^{-1}z_2)\right)
%\right]=0.
\end{eqnarray}
The basic operators
$:\exp\left(b_\pm^{i,j}(z)\pm (b+c)^{i,j}(q^{\mp 1}z)\right):$
create the delta-function $\delta(z)$ and
play important roles in constructions
of the bosonic operators $X_i^\pm(z)$ $(i \neq N)$ 
that satisfy
\begin{eqnarray}
~[X_i^+(z_1),X_j^-(z_2)]
=\frac{\delta_{i,j}}{(q-q^{-1})z_1z_2}
\left(
\delta(q^{k}z_2/z_1)\Psi_i^+(q^{\frac{k}{2}}z_2)-
\delta(q^{-k}z_2/z_1)\Psi_i^-(q^{-\frac{k}{2}}z_2)
\right).\nonumber
\end{eqnarray}
Multiplying and adding proper 
operators to these basic operators
(\ref{def:basic}),
we construct 
the free field realization.
For this purpose,
the following replacement
from the Heisenberg realization of $U_q(sl(N|1))$ 
to the free field realization of the affine 
$U_q(\widehat{sl}(N|1))$ gives useful information.

\subsection{Replacement}

In this appendix 
we explain how to find the free field realization
of the affine superalgebra $U_q(\widehat{sl}(N|1))$
from the Heisenberg realization of $U_q(sl(N|1))$.
We make the following replacement with suitable argument.
\begin{eqnarray}
\vartheta_{i,j}&\to&-b_\pm^{i,j}(z)/{\rm log}q~~~~~
(1\leqq i<j \leqq N+1),\\
~[\vartheta_{i,j}]_q&\to&
\left\{
\begin{array}{cc}
\frac{\displaystyle
\exp\left(\pm b_+^{i,j}(z)\right)-
\exp\left(\pm b_-^{i,j}(z)\right)}{
\displaystyle
(q-q^{-1})z}
&~~~(j\neq N+1),\\
1&~~~(j=N+1).
\end{array}
\right.
\\
x_{i,j}
&\to&
\left\{
\begin{array}{cc}
:\exp\left((b+c)^{i,j}(z)\right):
&~~(j\neq N+1),\\
:\exp\left(-b^{i,j}(z)\right):
~{\rm or}~
:\exp\left(-b_\pm^{i,j}(q^{\pm 1}z)-b^{i,j}(z)\right):
&~~(j=N+1).
\end{array}
\right.
\label{def:replacement}\\
\lambda_i
&\to&
a_\pm^i(z)/{\rm log}q~~~~~(1\leqq i \leqq N),\\
~[\lambda_i]_q
&\to&
\frac{\exp\left(\pm a_+^i(z)\right)-
\exp\left(\pm a_-^i(z)\right)}{
\displaystyle (q-q^{-1})z}~~~~~(1\leqq i \leqq N).
\end{eqnarray}

~\\
Taking the basic operators (\ref{def:basic}) into account,
we gave this rule of the replacement.

From the above replacement,
$H_i$ of the Heisenberg realization
(\ref{def:Heisenberg}) is replaced as following.
\begin{eqnarray}
q^{H_i}
\to
\left\{
\begin{array}{cc}
\exp\left(a_\pm^i(z)
+\sum_{l=1}^i(b_\pm^{l,i+1}(z)-b_\pm^{l,i}(z))
+\sum_{l=i+1}^N (b_\pm^{i,l}(z)-b_\pm^{i+1,l}(z))\right)&
~(1\leqq i \leqq N-1),\\
\exp\left(
a_\pm^N(z)-\sum_{l=1}^{N-1}
(b_\pm^{l,N}(z)+b_\pm^{l,N+1}(z))
\right)&~(i=N).
\end{array}
\right.\nonumber\\
\label{replacement:h}
\end{eqnarray}
There exist small gaps between
the above operators (\ref{replacement:h}) and
the free field realizations $\Psi_i^\pm(z)$ (\ref{boson5}), 
(\ref{boson6}).
In order to make
the operators (\ref{replacement:h}) satisfy
the defining relations of $U_q(\widehat{sl}(N|1))$,
we have to impose $q$-shift to variable $z$ of 
the operators $a^i_\pm(z)$,
$b_\pm^{i,j}(z)$. 
For instance, we have to replace 
$a^i_\pm(z) \to a^i_\pm(q^{\pm \frac{k+N-1}{2}}z)$.
Bridging the gap by the $q$-shift,
we have the free field realizations
$\Psi_i^\pm(q^{\pm\frac{k}{2}}z)$ (\ref{boson5}),
(\ref{boson6}) from $q^{H_i}$.
\begin{eqnarray}
q^{H_i} \to \Psi_i^\pm (q^{\pm \frac{k}{2}}z)
~~~(1\leqq i \leqq N).
\end{eqnarray}

The structure of non-superalgebra $U_q(\widehat{sl}(N))$ exists inside
the superalgebra $U_q(\widehat{sl}(N|1))$.
Hence the free field realizations of the currents 
$X_i^\pm(z)$ $(i \neq N)$
for $U_q(\widehat{sl}(N|1))$ 
are quite similar as those for
$U_q(\widehat{sl}(N))$.
Let us focus our attention on
the fermionic operators $X_N^\pm(z)$
that is new for the superalgebra.
Let us consider $E_N=\sum_{j=1}^N E_{N,j}$
of the Heisenberg realization (\ref{def:Heisenberg}).
From the above replacement, we have
\begin{eqnarray}
E_{N,j}&\to&
:\exp
\left((b+c)^{j,N}(z)+b^{j,N+1}(z)-
\sum_{l=1}^{j-1}(b_+^{l,N}(z)+b_+^{l,N+1}(z))\right):.
\end{eqnarray}
There exists an ambiguity of the replacement of
$x_{j,N+1}$ in (\ref{def:replacement}).
Here we have chose the replacement
$x_{j,N+1} \to :\exp\left(-b^{j,N+1}(z)\right):$
$(1\leqq j \leqq N)$.
Imposing proper $q$-shift to the variable $z$ of 
the operators $(b+c)^{j,N}(z)$, $b^{j,N+1}(z)$,
$b_\pm^{i,j}(z)$, we have
the free field realizations
$X_{N,j}^+(z)$ in (\ref{boson9}).
\begin{eqnarray}
E_{N,j} \to X_{N,j}^+(z)~~~(1\leqq j \leqq N).
\end{eqnarray}
Let us consider $F_N=\sum_{j=1}^N F_{N,j}$
of the Heisenberg realization (\ref{def:Heisenberg}).
From the above replacement we have
\begin{eqnarray}
&&F_{N,j}
\to
\frac{1}{(q-q^{-1})z}\times
\nonumber\\
&\times&
\left\{
\begin{array}{cc}
\begin{array}{c}
:\exp
\left(
-a_-^N(z)-b^{j,N+1}(z)-(b+c)^{j,N}(z)+
\sum_{l=j+1}^{N-1}(b_-^{l,N+1}(z)-b_-^{l,N}(z))
\right)\\
\times
\left(
\exp\left(-b_+^{j,N}(z)-b_+^{j,N+1}(z)\right)-
\exp\left(-b_-^{j,N}(z)-b_-^{j,N+1}(z)\right)
\right):
\end{array}
&~(j\neq N),\\
:\exp\left(-b^{N,N+1}(z)\right)
\left(
\exp\left(a_+^N(z)\right)-\exp\left(a_-^N(z)\right)
\right):&~~(j=N).
\end{array}
\right.
\nonumber\\
\end{eqnarray}
There exists an ambiguity of the replacement of
$x_{j,N+1}$ in (\ref{def:replacement}).
Here we have chose
the replacement
$x_{j,N+1} \to 
:\exp\left(b_\pm^{j,N+1}(q^{\mp 1}z)-b^{j,N+1}(z)\right):$
~$(1\leqq j \leqq N-1)$ and
$x_{N,N+1} \to 
:\exp\left(-b^{N,N+1}(z)\right):$.
Imposing proper $q$-shift to the variable $z$ of 
the operators $(b+c)^{j,N}(z)$, $b^{j,N+1}(z)$,
$b_\pm^{i,j}(z)$, $a_-^N(z)$, we have
the free field realizations
$X_{N,2j-1}^-(z), X_{N,2j}^-(z)$ in (\ref{boson17}),
(\ref{boson18}), (\ref{boson19}) and (\ref{boson20}).
\begin{eqnarray}
F_{N,j} \to \frac{-1}{(q-q^{-1})z}(
X_{N,2j-1}^-(z)-X_{N,2j}^-(z))~~~(1\leqq j \leqq N).
\end{eqnarray}
Replacements for bosonic operators 
$X_j^\pm(z)$, $(j \neq N)$ 
have already appeared in 
$U_q(\widehat{sl}(N))$
\cite{Awata-Odake-Shiraishi1}.
We explained
details of the replacement
for the fermionic operator $X_N^\pm(z)$,
which is new for the superalgebra.

\section{Normal Orderings}
\label{appendixB}

In this appendix we summarize useful relations.\\
For $1\leqq j \leqq N$ we have
\begin{eqnarray}
\{X_{N,j}^+(z_1),X_{N,2j-1}^-(z_2)\}
&=&
\frac{1}{q^{j-1}z_1}\delta(q^{-k-2j+2}z_2/z_1)\nonumber\\
&\times&
:\exp\left(a_-^N(q^{-\frac{k+N-1}{2}}z_2)
-\sum_{l=1}^{j-1}(
b_+^{l,N}(q^{-k-2j+l+2}z_2)+b_+^{l,N+1}(q^{-k-2j+l+2}z_2))\right.
\nonumber\\
&&\left.
-\sum_{l=j}^{N-1}(b_-^{l,N}(q^{-k-l}z_2)+b_-^{l,N+1}(q^{-k-l}z_2))
\right):.
\label{eqn:a1}
\end{eqnarray}
Especially for $j=1$ we have
\begin{eqnarray}
&&\left\{X_{N,1}^+(z_1),X_{N,1}^-(z_2)\right\}=\frac{1}{z_1}
\delta(q^{-k}z_2/z_1)\Psi_N^-(q^{-\frac{k}{2}}z_2).
\label{eqn:a2}
\end{eqnarray}
For $1\leqq j \leqq N-1$ we have
\begin{eqnarray}
\{X_{N,j}^+(z_1),X_{N,2j}^-(z_2)\}
&=&\frac{1}{q^{j-1}z_1}
\delta(q^{-k-2j}z_2/z_1)\nonumber\\
&\times&
:\exp\left(
a_-^N(q^{-\frac{k+N-1}{2}}z_2)
-\sum_{l=1}^{j}(
b_+^{l,N}(q^{-k-2j+l}z_2)+b_+^{l,N+1}(q^{-k-2j+l}z_2))\right.\nonumber\\
&&
\left.-\sum_{l=j+1}^{N-1}(b_-^{l,N}(q^{-k-l}z_2)+b_-^{l,N+1}(q^{-k-l}z_2))
\right):,
\label{eqn:a3}
\\
\{X_{N,N}^+(z_1),X_{N,2N}^-(z_2)\}
&=&\frac{1}{q^{N-1}z_1}
\delta(q^kz_2/z_1)\Psi_N^+(q^{\frac{k}{2}}z_2).
\label{eqn:a4}
\end{eqnarray}
Other anti-commutators relations 
$\left\{X_{N,i}^+(z_1),X_{N,j}^-(z_2)\right\}$ vanish.
\begin{eqnarray}
\left\{X_{N,i}^+(z_1),X_{N,j}^-(z_2)\right\}=0~~~
{\rm for}~j \neq 2i-1, 2i.
\label{eqn:a5}
\end{eqnarray}
For $1\leqq j \leqq N-2$ we have
\begin{eqnarray}
~[X_{N,j+1}^+(z_1),X_{j,2N-3}^-(z_2)]
&=&(q-q^{-1})\delta(q^{k+N-j}z_2/z_1)\nonumber\\
&\times&
:\exp\left(a_+^{j}(q^{\frac{k+N-1}{2}}z_2)
-b_+^{j+1,N+1}(q^{k+N-1}z_2)\right.\nonumber\\
&&\left.+b^{j+1,N+1}(q^{k+N}z_2)
+(b+c)^{j,N}(q^{k+N-1}z_2)\right.\nonumber\\
&&\left.-\sum_{l=1}^{j-1}
(b_+^{l,N}(q^{k+N-j+l}z_2)+b_+^{l,N+1}(q^{k+N-j+l}z_2))\right):.
\label{eqn:a6}
\end{eqnarray}
We have
\begin{eqnarray}
~[X_{N,N}^+(z_1),X_{N-1,2N-2}^-(z_2)]
&=&(q-q^{-1})
\delta(q^{k+1}z_2/z_1)\nonumber\\
&\times&
:\exp\left(
a_+^{N-1}(q^{\frac{k+N-1}{2}}z_2)
-b_+^{N,N+1}(q^{k+N-1}z_2)\right.\nonumber\\
&&\left.+b^{N,N+1}(q^{k+N}z_2)
+(b+c)^{N-1,N}(q^{k+N-1}z_2)\right.\nonumber\\
&&\left.
-\sum_{l=1}^{N-2}
(b_+^{l,N}(q^{k+l+1}z_2)+b_+^{l,N+1}(q^{k+l+1}z_2))\right):.
\label{eqn:a7}
\end{eqnarray}
For $1\leqq j \leqq N-1$ we have
\begin{eqnarray}
~[X_{N,j}^+(z_1),X_{j,2N-1}^-(z_2)]
&=&\frac{1}{q^{j-1}z_1}\delta(q^{k+N-j}z_2/z_1)\nonumber\\
&\times&
:\exp\left(a_+^{j}(q^{\frac{k+N-1}{2}}z_2)
-b_+^{j+1,N+1}(q^{k+N-1}z_2)\right.\nonumber\\
&&\left.+b^{j+1,N+1}(q^{k+N}z_2)
+(b+c)^{j,N}(q^{k+N-1}z_2)\right.\nonumber\\
&&\left.-\sum_{l=1}^{j-1}
(b_+^{l,N}(q^{k+N-j+l}z_2)+b_+^{l,N+1}(q^{k+N-j+l}z_2))\right):.
\label{eqn:a8}
\end{eqnarray}
Other commutation relations 
$\left[X_{N,i}^+(z_1),X_{l,j}^-(z_2)\right]$
vanish.
\begin{eqnarray}
\left[X_{N,i}^+(z_1),X_{j,l}^-(z_2)\right]=0~~~~
{\rm for}~~(i,j,l)\neq
\left\{
\begin{array}{cc}
(j,j,2N-1)&~~(1\leqq j \leqq N-1),\\
(j+1,j,2N-3)&~~(1\leqq j \leqq N-2),\\
(N,N-1,2N-2)&.
\end{array}
\right.
\label{eqn:a9}
\end{eqnarray}
For $1\leqq i \leqq N-1$ we have
\begin{eqnarray}
&&[X_{i,2}^+(z_1),X_{i,1}^-(z_2)]=(q-q^{-1})
\delta(q^{-k}z_2/z_1)\Psi_i^-(q^{-\frac{k}{2}}z_2),
\label{eqn:a10}\\
&&[X_{i,2i-1}^+(z_1),X_{i,2i}^-(z_2)]=-(q-q^{-1})
\delta(q^{k}z_2/z_1)\Psi_i^+(q^{\frac{k}{2}}z_2).
\label{eqn:a11}
\end{eqnarray}

~\\

\end{appendix}

\end{document}